\documentclass [twocolumn,showpacs,prb,floatfix,amsmath,amssymb] {revtex4}

\usepackage{graphicx}

\begin{document}

\title{A Millikelvin Scanned Probe for Measurement of Nanostructures}

\author{K. R. Brown}
\email{krbrown@physics.umd.edu}
\author{L. Sun}
\author{B. E. Kane}
\affiliation{Laboratory for Physical Sciences, 8050 Greenmead Drive,
  College Park, Maryland, 20740}
\begin{abstract}
We demonstrate a scanning force microscope, based upon a quartz tuning
fork, that operates below 100 mK and in magnetic fields up to 6 T. The
microscope has a conducting tip for electrical probing of
nanostructures of interest, and it incorporates a low noise cryogenic
amplifier to measure both the vibrations of the tuning fork and the
electrical signals from the nanostructures. At millikelvin
temperatures the imaging resolution is below 1 $\mu$m in a 22 $\mu$m x
22 $\mu$m range, and a coarse motion provides translations of a few
mm.  This scanned probe is useful for high bandwidth measurement of
many high impedance nanostructures on a single sample. We show data
locating an SET within an array and measure its coulomb blockade
with a sensitivity of $2.6 \cdot 10^{-5}$ $e/\sqrt{Hz}$.
\end{abstract}


\maketitle

\section{Introduction}
A variety of scanned probe techniques has been adapted or developed to
study nanostructures and transport at cryogenic temperatures,
including atomic force microscopy,\cite{pelekhov,rychen} scanned gate
microscopy,\cite{eriksson} scanning capacitance
microscopy,\cite{tessmer} and Kelvin probe microscopy.\cite{vancura}
Crook et al. recently even reported a technique for nanolithography
using a scanned probe at dilution refrigerator
temperatures.\cite{crook} Here we describe a cryogenic scanning force
microscope (SFM) operating at millikelvin temperatures and in high
magnetic fields. The SFM was motivated by a desire to locate and probe
large numbers of single electron transistors (SET's), in order to
measure single donors in silicon and to observe single charge
motion.\cite{kane2} SET's and quantum point contacts are the most
sensitive devices yet discovered for measuring electric charge, with a
sensitivity near the theoretical quantum limit of $10^{-6}$
e$/\sqrt{Hz}$. Nevertheless, the high impedance of these devices and
their required operating temperature in the millikelvin regime has
generally restricted their use to low frequencies in a dilution
refrigerator. One solution to this bandwidth problem is the
RF-SET.\cite{schoelkopf} Another solution is a preamplifier close to
the SET or point contact that minimizes parasitic
capacitances.\cite{visscher, pohlen} Although frequency domain
multiplexing RF-SET's can be used with small numbers of
devices,\cite{stevenson} in general both of these solutions suffer
from the limitation that measuring a new device requires a full warmup
and cooldown cycle of the refrigerator. Our millikelvin scanned probe
circumvents this problem, effectively acting as a multiplexer. The
imaging capability of the SFM locates a particular SET within an
array, and a conducting probe tip provides an electrical connection
between that SET and a cryogenic preamplifier for measurements.
\section{Experimental Setup}
\begin{figure}
\includegraphics [width=3.375in] {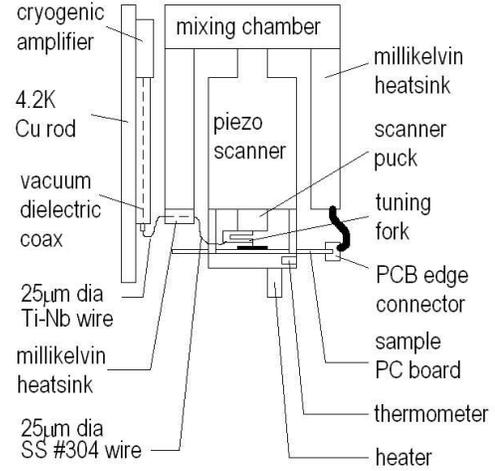}
\caption{Physical layout of the scanning force microscope. The piezo
  scanner is attached to the mixing chamber of the dilution
  refrigerator, and the sample is mounted on a printed circuit board
  suspended below the scanner. All the leads to the sample are
  heatsunk to the mixing chamber. The cryogenic amplifier is attached
  to a 4.2 K finger that comes down alongside the scanner, and the
  tuning fork is connected to the amplifier via 25 $\mu$m diameter
  wires. The vertical distance between the tuning fork and the
  amplifier is approximately 20 cm, placing the amplifier outside any
  applied magnetic field. The entire setup is at cryogenic UHV
  pressure inside the dilution refrigerator vacuum can.}
\label{fig:diagram}
\end{figure}

Fig.~\ref{fig:diagram} shows the physical layout of the microscope.
The force sensor is a quartz tuning fork (Raltron R38-32.768-12.5)
with a sharpened W tip glued to one of its tines.\cite{raltron}
The sub-pW power dissipation of the fork has no measureable effect
on the base temperature of our refrigerator, and the laser-free
piezoelectric detection scheme avoids problems with light sensitive
samples.  The scanner is a commercial cryogenic STM scan head
(Omicron's CryoSXM\cite{cryosxm}) modified for SFM and mounted on the
baseplate of an Oxford Kelvinox 100 dilution refrigerator. The
microscope incorporates a cryogenic amplifier for maximum sensitivity
and bandwidth. We use a phase-locked loop (Nanosurf
easyPLL\cite{easypll}) to excite the fork.

Piezoelectric quartz tuning forks have been used as force sensors in
scanning probe microscopy for several years now. Mechanical and
electrical models\cite{grober,karrai,rychen2,rychen3} predict that
forces acting on a fork change its resonant frequency and
Q. Frictional forces change Q while conservative forces change the
frequency. Our microscope uses the frequency shift as the force
dependent signal, varying the tip height to maintain a constant shift
while rastering over the sample surface.

\begin{figure}
\includegraphics [width=3.375in] {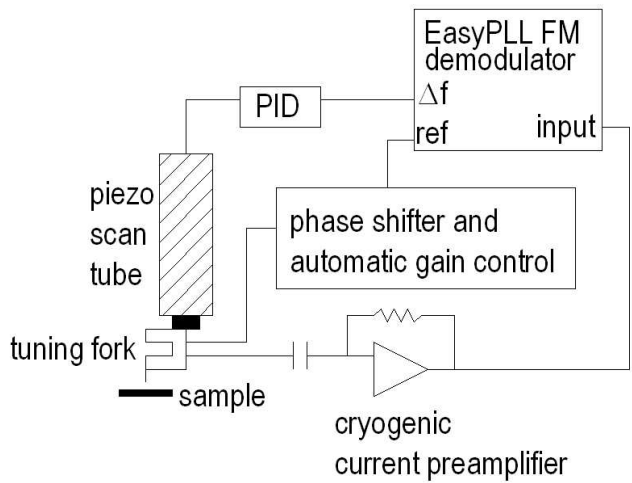}
\caption{Block diagram of the cryogenic scanning force microscope. PID
  circuitry controls the scan piezo to maintain a constant frequency
  shift of the tuning fork. The fork is driven by the reference output
  of the easyPLL with a suitable phase shift and automatic gain
  control. The response is measured with a cryogenic amplifier and
  sent to the input of the easyPLL. The cryogenic amplifier is
  simultaneously connected to the metallic tip on the tuning fork and
  to the tuning fork electrode.}
\label{fig:schematic}
\end{figure}

Fig.~\ref{fig:schematic} shows a block diagram of the microscope's
operation.
The tuning fork is excited with an AC potential applied to one of its
electrodes. The other electrode is held at virtual ground by the input
of a cryogenic transimpedance amplifier, discussed below. We use a
bridge circuit\cite{grober} to eliminate the effects of stray
capacitance between the two electrodes. The current flowing between
the two electrodes, as measured by the transimpedance amplifier, is a
simple but effective way to measure the motion of the fork.

We use a phase-locked loop to monitor the frequency of the fork and to
insure that it is always driven on resonance.\cite{albrecht} This PLL
has a nominal sensitivity of 5 mHz in a 1 kHz bandwidth, but we obtain
a sensitivity less than this in practice, typically a few tens of
mHz. This sensitivity is limited primarily by white noise from the
cryogenic amplifier near the resonant frequency.

The scanner was originally designed designed for STM at 4.2 K in a
helium exchange gas environment. We have detached the scan head from
its original mounting rod and attached it to the base plate of the
dilution refrigerator. The high-voltage signals for piezo control are
carried on twisted pairs from room temperature to 4.2 K and on
superconducting twisted pairs from 4.2 K to the mixing chamber,
thereby minimizing Johnson heating of the refrigerator while
scanning. In contrast with Omicron's intended configuration, we scan
the tuning fork with the sample fixed. This allows us to connect
multiple leads to the sample and to control its temperature more
easily.

The maximum $\pm$135 V scan voltage provided by the Omicron
electronics gives a range of 35 $\mu$m x 35 $\mu$m at room
temperature. This decreases to 7 $\mu$m x 7 $\mu$m at 4.2 K and below,
a range that is frequently too small to be useful. The z-range
similarly decreases from 2.7 $\mu$m to 0.5 $\mu$m. To increase the
scan range at low temperatures we have added an additional
high-voltage amplifier (RHK HVA-900\cite{rhk}) that provides scan
voltages up to $\pm$450 V, yielding a maximum scan range of 22 $\mu$m
x 22 $\mu$m and a maximum z-range of 1.8 $\mu$m below 4.2 K. The
scanner also provides a slip-stick based coarse motion both in the x-y
plane and in the z-direction. Motions of a few mm in x and y and up to
a few cm in z is possible. However, we have at times found coarse
motion in the x-y plane to be unreliable.
\section{Cryogenic Amplifier Electronics}
The microscope requires a transimpedance amplifier to measure the
current through the tuning fork while scanning. The same amplifier
measures the current through an SET while probing. To decrease the
unavoidable parasitic capacitances in these measurements we have
constructed a cryogenic amplifier and placed it near the sample on a
copper rod heat sunk to 4.2 K. As an added benefit the cryogenically
cooled current-sensing resistor has significantly lower Johnson noise
than it has at room temperature. The amplifier is based upon a
low-noise silicon JFET (Moxtek MX120,\cite{moxtek} input-referred
voltage noise $\sim$ 3 nV$/\sqrt{Hz}$, 1/f knee below 1 kHz) that
stops working below $\sim 60$ K. In order to maintain most of the
circuit at 4.2 K while allowing the JFET to heat above 60 K we have
designed the circuit with two printed circuit boards. The lower board
contains most of the circuit elements and is held at 4.2 K. The upper
board containing the JFET is mounted to the lower one with $1/16$''
diameter nylon standoffs. All connections between the two boards are
made with 25 $\mu$m diameter stainless steel wire to minimize thermal
loading. The amplifier dissipates $\sim$ 20 mW during operation and
raises the base temperature of the refrigerator by less than 0.5 mK to
about 15 mK.  To achieve such a small thermal load on the mixing
chamber it is crucial that the copper box enclosing the amplifier is
leak-tight, light-tight, and well heat sunk to the 4.2 K bath. A
charcoal sorb inside the box captures residual gas that otherwise
would prevent the JFET from heating properly.

The input to the amplifier is capacitively coupled via a 4.7 nF
capacitor and a coaxial cable to one of the tuning fork electrodes,
and via the metallization on the tuning fork to the conducting probe
tip. We capacitively couple the signal to avoid problems with DC
offsets. For thermalization at the mixing chamber the coax has a 2 cm
length of Stycast 2850FT as its dielectric, which contributes $\sim$
18 pF stray capacitance to ground. The remainder of the coax has a
vacuum dielectric to minimize its capacitance, so that the total stray
input capacitance, including the gate capacitance of the JFET, is only
$\sim$ 35 pF. This means that for a 50 k$\Omega$ SET the bandwidth of
measurement is limited to $\sim 100$ kHz. Higher speeds are possible
only at the expense of further noise.
\section{Scan Modes}
For the purposes of locating micron-sized pads and contacting a large
number of them in a reasonable period of time, scan speed rather than
scan resolution is the critical factor. We would ideally like to scan
the entire 22 $\mu$m range of the scanner in a few minutes. We have
investigated three different force regimes or scan modes: a
short-range repulsive force mode, a weaker attractive force mode, and
a long-range electrostatic force mode. The repulsive mode,
corresponding to positive frequency shifts of a few Hz, yields the
highest resolution images but is the least useful for our
purposes. The very small tip-sample separation and highly nonlinear
frequency vs. separation dependence make optimization of the feedback
loop difficult and leads to frequent tip crashes at reasonable speeds.
Faster scanning is possible with a weaker attractive force,
corresponding to negative frequency shifts of a few hundred
mHz. However, with this scan mode we were only successful using
smaller, more sensitive forks that have other disadvantages as
explained below in Section~\ref{section:experimental challenges}. The
fastest scanning is obtained by charging up the tip through a diode,
so that it feels a strong attractive force from the image charges in
the sample.\cite{seo} The resulting force is of sufficient range to
have easily measured effects even with tip-sample separations greater
than 50 nm. This scan mode even works on insulating substrates
(e.g. SiO$_2$) provided that the dielectric constant of the substrate
differs from that of the vacuum, and yields scanning speeds of 20
$\mu$m/s or more.
\section{Samples}
\begin{figure}
\includegraphics [width=3.375in] {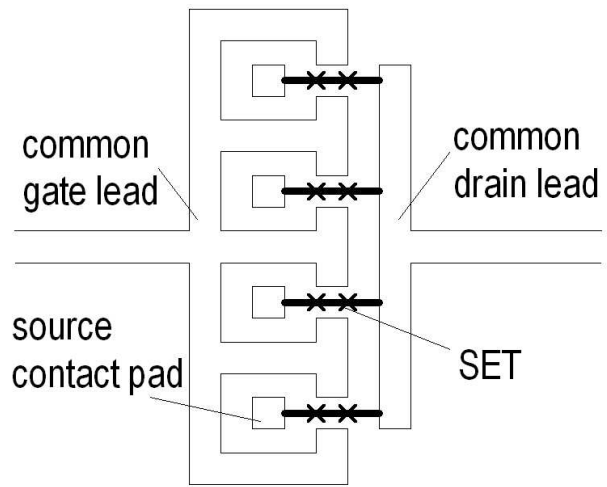}
\caption{Schematic of an SET array showing common gate and drain
  leads. Each SET has a contact pad attached to its source terminal so
  that it may be measured with the scanned probe.}
\label{fig:array}
\end{figure}

Our samples consist of a 100 $\mu$m x 100 $\mu$m array of
Al-AlO$_2$-Al SET's fabricated with standard e-beam lithography and
double-angle evaporation.\cite{dolan} Each SET has a small island
weakly coupled to source and drain leads via oxide tunnel barriers and
capacitively coupled to a gate. The drain-source current is strongly
dependent on gate voltage, provided that the resistance of the tunnel
junctions is $\agt h/e^{2} \approx 25.9$ k$\Omega$.\cite{fulton} A
schematic of the sample layout is shown in Fig.~\ref{fig:array}.
Within the array, all of the SET gate terminals are wired to a single
coaxial lead, and all of the drain terminals to another. The source
terminal of each SET is connected to its own 3 $\mu$m x 3 $\mu$m Pt
contact pad.

Sample fabrication proceeds through three layers of lithography. First
the common gate and drain leads as well as the source contact pads are
patterned with photolithography. A trilayer consisting of 30 \AA\
Ti, 120 \AA\ Pt, and 2500 \AA\ Au is deposited. Second,
we mask the Au bond pads with another layer of photolithography and
etch the Au away near the sample center to expose the Pt
layer. This leaves a thick Au layer for the bond pads but a thin
Pt layer for the SET array. The array metallization needs to be
both thin and oxide free. Thin metallization allows us to increase the
scan speed of the microscope. The metal surface must be oxide free to
insure low resistance contact between it and the Al SET
lithography, as well as to insure low resistance contact between the
contact pads and the microscope tip during probing. Third, we use
standard e-beam lithography to write the SET's and ash the sample to
remove any organics. The sample is attached with a spring clip to a
printed circuit board, and Au wires are attached to the
lithographically defined leads. It is important that the wires all
come off the sample in a single direction, away from where the scanned
probe will be, so as not to obstruct its motion.

Typical measurements proceed in two sequential steps. First we scan
the surface of the sample until we find an SET of interest. Then we
turn off the tip-sample feedback loop and establish electrical contact
between the probe tip and the Pt contact pad (lightly crash
the tip). An AC voltage (50 - 100 kHz) is applied to the drain
lead, and the current through the SET is monitored with the cryogenic
amplifier and a lock-in.

The only topographic information required for this measurement is the
location of the contact pads. These are large and separated from the
surrounding lithography by 2 $\mu$m on all sides. Therefore the
required resolution of the SFM is only about 1 $\mu$m. This fact is
important as the probe resolution tends to deteriorate over time with
repeated probing and cleaning.
\section{Results and Discussion}
\begin{figure}
\includegraphics [width=3.375in] {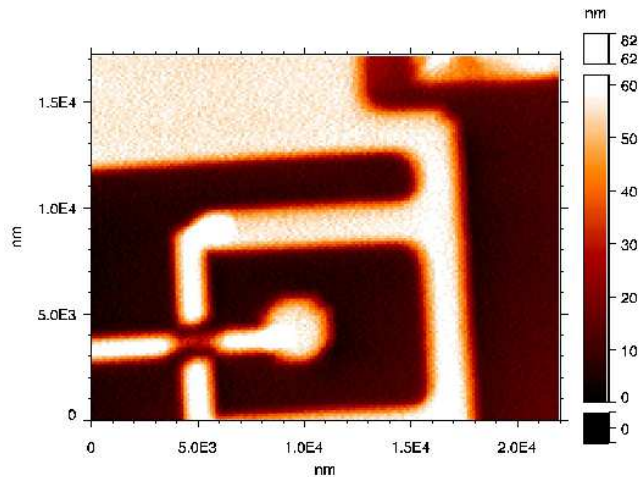}
\caption{SFM scan showing Pt and Al features lithographically
  defined on an oxidized silicon substrate. The scan was taken with
  V$_{tip}$ = -40 V and $\Delta$f = -0.5 Hz at a speed of 15
  $\mu$m/s. During scanning the sample temperature rose to 190 mK and
  the mixing chamber temperature rose to 35 mK. The 3 $\mu$m x 3
  $\mu$m square near the center of the image is the contact pad of the
  SET.}
\label{fig:scan}
\end{figure}

Fig.~\ref{fig:scan} is an image taken at millikelvin temperature
showing part of an SET array.
The image clearly demonstrates resolution well below 1 $\mu$m at a
scan speed of 15 $\mu$m/s and a scan range of 22 $\mu$m. Although we
observe no extra heating of the refrigerator while the scan piezo is held
motionless, there was enough heating during scanning to raise the
temperature of the sample from 15 mK to 190 mK. The amount of heating
depends strongly on the scan size and speed. In practice this heating
should not cause problems because the piezo is held motionless while
probing an SET.

\begin{figure}
\includegraphics [width=3.375in] {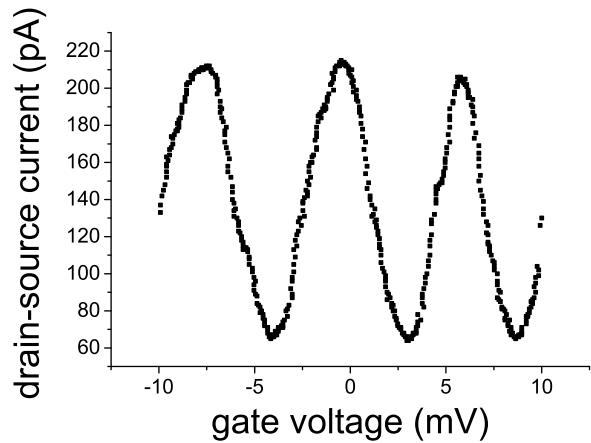}
\caption{Coulomb blockade oscillations of the current through a single
  electron transistor, as measured with the cryogenic scanned
  probe. The data were measured at 60 mK with B = 1 T and
  V$_{ds}$ = 100 $\mu$V rms.}
\label{fig:blockade}
\end{figure}

We have electrically probed several SET's on multiple samples.
Fig.~\ref{fig:blockade} shows characteristic coulomb blockade
oscillations.
A 100 kHz, 100 $\mu$V rms voltage was applied to the drain lead, and
the in-phase component of the resulting AC current was measured by a
lock-in. The SET was kept in the normal state by a 1 T magnetic
field. The current never drops all the way to zero on the modulation
curve because the 100 $\mu$V rms drain-source voltage is larger than
the coulomb blockade plateau. We observed flicker noise in these
measurements far in excess of what we would expect and are conducting
further experiments to determine the source.

We also made measurements of the charge sensitivity of the
SET/amplifier configuration. Applying a known amount of charge to the
gate of the SET and measuring the response at the output of the
lock-in calibrates the gain. Comparing this gain with the noise
measured at the output of the lock-in gives the charge sensitivity.
For our most recent devices with resistance $\sim$ 80 k$\Omega$ we
have measured a charge sensitivity of $2.6 \cdot 10^{-5}$
$e/\sqrt{Hz}$. This sensitivity is limited by the Johnson noise of our
feedback resistor (20 $nV/\sqrt{Hz}$) and by the input-referred
voltage noise of our amplifier.  The equivalent current noise values
are 20 $fA/\sqrt{Hz}$ and 38 $fA/\sqrt{Hz}$, comparable to the
intrinsic $\sim 10$ $fA/\sqrt{Hz}$ shot noise of the SET.
\section{Experimental Challenges} 
\label{section:experimental challenges}
Following the advice of Giessibl,\cite{giessibl} we at one point tried
to use smaller forks with an entire prong glued down. A smaller spring
constant should yield higher force sensitivity, and gluing down an
entire prong should give a Q independent of the mass of the tip, while
for a fork glued only at its base the tip mass breaks the symmetry of
the tines and leads to dissipation. However, we found that tuning
forks with an entire prong glued down exhibit a strong coupling with
the piezo scan tube. This manifests itself as seemingly random shifts
in the fork resonant frequency when the scan tube expands or
contracts, making scanning difficult or impossible. We observed no
similar shifts using forks glued only at the base, presumably because
in this case the vibration is decoupled from its support. In
constrast, a fork with its entire prong glued down incorporates its
support structure into the system, and we might expect that small
changes in the support structure could strongly affect its
vibrations. One way to avoid this problem would be to scan the sample
instead of the fork.

Another challenge is to establish low-resistance contacts reliably and
repeatably between the Pt contact pads and the probe tip. We have had
success with both PtIr and W tips and repeatably get contact
resistances less than 1 k$\Omega$, although to date we have been most
successful using W tips. We expected to have no difficulties with PtIr
tips due to the lack of native oxide but found that they were
contaminated easily.  Therefore we inevitably need to clean the tip
with field-emission into the sample before making the first contact
after cooldown. With the tip only $\sim$ 1 nm away from the sample
surface we apply successively more negative voltages until we see an
abrupt increase in the tip-sample current, indicating a clean tip. To
maintain a clean sample surface while cooling we take care
to maintain its temperature above that of the walls of the
vacuum space, insuring that any contaminants condense on the walls
instead.\cite{rychen}
\section{Conclusions and Prospects}
We have successfully demonstrated a scanning force microscope with
sub-micron resolution at dilution refrigerator temperatures and in
high magnetic fields. By imaging the sample to locate individual SET's
within a large array and making electrical contact to each SET in turn
with the conducting probe tip, we can measure large numbers of SET's
with low noise and high bandwidth during a single cooldown. While our
application is to arrays of SET's, the cryogenic scanned probe could
be applied to many other situations where it is necessary to measure
large numbers of high-impedance nanostructures.
\begin{acknowledgments}
We thank Thomas Ihn, Barry Barker, and Ray Phaneuf for helpful
discussions. This work was supported by the Advanced Research
and Development Activity and the National Security Agency.
\end{acknowledgments}
\bibliography{spm,sets,qc}
\end{document}